\documentclass[aps,prl,twocolumn,showpacs]{revtex4}
\usepackage{graphicx}
\begin{document}
\title{Localized Orbital Description of Electronic Structure}
\author{Joydeep Bhattacharjee and Umesh V Waghmare}
\affiliation{Theoretical Sciences Unit\\
Jawaharlal Nehru Centre for Advanced Scientific Research\\ Jakkur PO,
Bangalore 560 064 India}
\begin{abstract}
We present a simple and general method for construction of localized orbitals 
to describe electronic structure of extended periodic metals and insulators as 
well as confined systems. Spatial decay of these orbitals is found to
exhibit exponential behavior for insulators and power law for metals. 
While these orbitals provide a clear description of bonding, they can be 
also used to determine polarization of insulators. 
Within density functional theory, we illustrate applications of this method
to crystalline Aluminium, Copper, Silicon, PbTiO$_3$ and molecules such as 
ethane and diborane.
\end{abstract}
\maketitle
Localized orbitals, such as Wannier functions\cite{Wan}, have played an important
role in many aspects of quantum mechanical description of electrons in solids
and molecules.
They highlight atomic character of electrons\cite{kohn73} and elucidate the nature
of bonding. Their spatial localization makes them useful as basis functions for
efficient calculations of electronic structure that scale
linearly with system size\cite{siesta,galli,goedecker}. They are often useful
in the construction of lattice model Hamiltonians for structural
transitions\cite{rabe95} as well as strongly correlated systems\cite{ole}.
Wannier functions are intimately related to geometric phases of Bloch electron\cite{Zak, BW}
and electric polarization of insulators\cite{KSV,Resta}. Recent surge of interest
in Wannier functions is due to their use in modeling transport \cite{transport} and
electric field dependent properties of periodic solids.

Many schemes for construction of localized or Wannier orbitals have been presented over
the last four decades. To name a few,
early works of Foster and Boys\cite{boys} were based on maximizing 
the dipole moment matrix elements between orthonormal orbitals making 
their centroids maximally apart from each other,
followed by Reudenberg's \cite{reuden} approach of maximizing the orbital 
self-repulsion energies. Kohn presented a scheme\cite{kohn73} for arriving at 
exponentially localized real orthonormal WFs through a variational principle for
total energy. Maximally localized Wannier functions (MLWF)\cite{MV} were introduced 
by Marzari and Vanderbilt that minimize the variance of position operator $\vec r$, in
a given subspace of states. For example, orbitals maximally localized in a spatial direction 
$\alpha$ (hermaphrodite orbitals\cite{Sgi}) are eigenfunctions of position operator 
$r_{\alpha}$ in the given subspace. As different components of $\vec r$ do not commute when
projected into a subspace, MLWFs are obtained through numerical minimization of 
variance of $\vec r$ used as a measure of localization. MLWFs have been
generalized to the cases of entangled bands\cite{Souza} and to 
atom-centered orbitals\cite{kosov}. 

WFs are Fourier transform of Bloch functions: $W(\mathbf{r})=\int d\mathbf{k} e^{i 
\mathbf{k}\cdot\mathbf{r}} \psi_{\mathbf{k}}( \mathbf{r})$. Due to freedom in 
{\bf k}-dependent phase factor accompanying $\psi_{\mathbf{k}}$, WFs are non-unique.
Smoothness of these phases determine localization properties of WFs. 
MLWFs in 1-dimension can be obtained simply ({\it with out} a variational calculation) 
through construction of Bloch states that are smooth (generated using parallel transport 
along {\bf k}) and periodic in $k-$space\cite{BW}. Generalization of this idea to three 
dimensions is not readily possible, as the smoothness and 
periodicity of Bloch states as a function of $k_{\alpha}$ achieved through
parallel transport along one direction is disturbed by that along another direction in
$k-$space. 
Here, we provide a solution by connecting Bloch states at different {\bf k}'s by parallel
transport along paths that run {\it outside} the {\bf k}-space.
Resulting Bloch functions are periodic and optimally smooth in all directions in $k-$space, 
whose Fourier transform
yields well localized Wannier functions. It is efficient and simple because it
avoids a variational calculation and issues of local minima. Its versatility is demonstrated
through applications to insulators, metals and molecules.

In the first step of this method, we develop an {\it auxiliary subspace} of highly localized 
orthonormal orbitals
with desired symmetry properties described by (a) center of the orbital (Wyckoff site) 
$\left\{\tau_\kappa\right\}$
and (b) irreducible representation (irrep) of its site symmetry group according to which it 
transforms\cite{kohn73,des}. This choice can be guided through symmetry analysis of Bloch 
states\cite{Zak, rabe95, smirnov} in the {\it physical subspace} of occupied states. For simplicity,
we use Gaussian form for the radial part and a spherical harmonic corresponding to the 
irrep of localized orbitals: 
\begin{equation}
\Psi_\mu(\mathbf{R,r}) = (\alpha/2\pi)^{3/2} e^{-\alpha |\mathbf{r-\tau_\kappa-R}|^2}Y_{lm}
(\widehat{r-\tau_\kappa-R})
\label{orb}
\end{equation}
where $\alpha$ determines the width of the Gaussian, {\bf R} is a direct space lattice
vector and $\mu$ is an orbital index. $\alpha$ is chosen to be large enough to keep 
orbitals in neighbouring unit cells orthogonal.
Cell periodic part $v_{\mu \mathbf{k}}$  of Bloch functions spanning the auxiliary subspace
is given by Fourier transform:
\begin{equation}
\langle \mathbf{r}| v_{\mu  \mathbf{k}}\rangle = \sum_R e^{i\mathbf{k}. \mathbf{(R-r)}} 
\Psi_\mu(\mathbf{R},r)
\end{equation}
We note that \{ $\langle \mathbf{r}| v_{\mu  \mathbf{k}}\rangle$ \} are
smooth as a function of {\bf k} and satisfy 
$|v_{\mu \mathbf{k}}\rangle = |v_{\mu \mathbf{k+G}}>$, {\bf G} being a reciprocal lattice vector.

In the second step, we perform a unitary transformation of Bloch states in the physical 
subspace (eg. energy eigenfunctions of occupied states) such that
the open path\cite{SB} non-abelian geometric phases between states in physical and auxiliary 
subspaces at fixed {\bf k} vanish. Overlap matrix 
$S^{\mathbf{k}}_{\mu n} = \langle v_{\mu \mathbf{k}} | u_{n \mathbf{k}}  \rangle$,
where $|u_{n \mathbf{k}}\rangle$ is cell periodic part of energy an eigen state, relates to
geometric phase matrix $\Gamma_k$ through:
\begin{equation}
S^{\mathbf{k}} = R_{\mathbf{k}} e^{i \Gamma_\mathbf{k}},
\label{otog}
\end{equation}
where $R$ is a positive definite Hermitian matrix. Determination of $R$ and $\Gamma_{\mathbf{k}}$
is accomplished using singular value decomposition of $S$:
\begin{equation}
S^{\mathbf{k}} = U_{\mathbf{k}} \Sigma_{\mathbf{k}} V_{\mathbf{k}}^\dagger \\
\label{rotn}
\end{equation}
where $\Sigma_{\mathbf{k}}$ is a diagonal matrix with non-negative singular values in its diagonal. 
Vanishing of singular value(s) at some $\mathbf{k}$ signals a non-optimal choice of symmetries of
localized orbitals in the auxiliary subspace and the need for correction.
Using Eqn.\ref{rotn}, $R_{\mathbf{k}}=U_{\mathbf{k}}\Sigma_{\mathbf{k}} U_{\mathbf{k}}^\dagger$ 
and $e^{i \Gamma_\mathbf{k}}=U_{\mathbf{k}}V_{\mathbf{k}}^\dagger$. 
A unitary transformation on the $\left\{ u_{n \mathbf{k}}\right\}$ given by a 
matrix $M_{\mathbf{k}} = (U_{\mathbf{k}}V_{\mathbf{k}}^\dagger)^\ast$ transforms the overlap 
matrix $S^{\mathbf{k}}$ to a 
hermitian form, 
$$ \tilde{S}^{\mathbf{k}} = S^{\mathbf{k}}(V_{\mathbf{k}}U_{\mathbf{k}}^\dagger) = 
U_{\mathbf{k}} \Sigma_{\mathbf{k}} U_{\mathbf{k}}^\dagger, $$
corresponding to vanishing geometric phases.

\begin{figure}[t]
\includegraphics[scale=0.40]{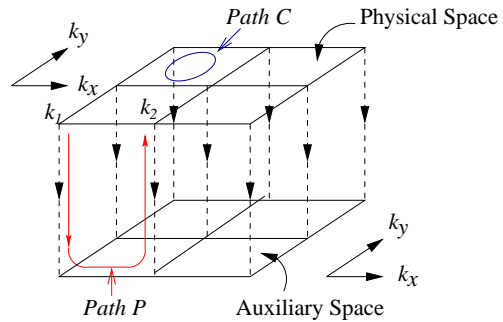} 
\caption{
Two horizontal planes are used to display physical and auxiliary subspaces, in which the 
member functions are parametrized with $k_x$ and $k_y$. Dashed lines indicate paths connecting
states at a given {\bf k} in physical subspace with those in the auxiliary subspace.}
\label{path}
\end{figure}

A sketch in terms of paths that connect physical and auxiliary subspaces (shown in Fig \ref{path})
enables understanding of our method. Closed paths parametrized by {\bf k},
such as $C$, in the physical subspace
can have nontrivial geometric phases which forbid construction of smooth functions by
parallel transport along {\bf k} within the physical subspace. In the present method,
we make use of paths, labeled as $P$, to connect states at two {\bf k}'s in the physical space. For 
$\mathbf{ k}=$ constant segments of path $P$ (dashed lines), geometric phases 
are made to vanish. Secondly, geometric phases along any closed path parametrized by {\bf k} in 
the auxiliary subspace vanish. Hence, Bloch functions in the physical subspace transformed 
with $M_{\mathbf{k}}$ have a single valued and periodic phase as a function of {\bf k}.

We include occupation numbers $f_{nk}$ while transforming energy eigenfunctions and obtain
\begin{eqnarray}
| \tilde{ u}_{\mu \mathbf{k}} \rangle = \sum_j M_{\mu j} | u_{j \mathbf{k}} \rangle
f_{nk}^{1/2},
\label{newfns}
\end{eqnarray}
which are periodic in {\bf k} by construction and the random phases 
accompanying $\left\{ u_{n \mathbf{k}}\right\}$ are filtered out in this procedure. 
Localized orbitals are obtained by Fourier transforming 
$\left\{ \tilde{ u}_{\mu \mathbf{k}}(\mathbf{r})\right\}$:
\begin{eqnarray}
<\mathbf{r}|\Phi_\mu (\mathbf{R}) \rangle= \frac{\Omega}{(2\pi)^3}\sum_{\mathbf{k}}e^{i\mathbf{(k-R).r}}<\mathbf{r}| \tilde{ u}_{\mu \mathbf{k}} \rangle
\label{phi}
\end{eqnarray}
where $\Omega$ is the unit cell volume. Localization of $|\Phi_\mu (\mathbf{R})\rangle$
is determined by the ``smoothness" of 
$\left\{ \tilde{ u}_{\mu \mathbf{k}}(\mathbf{r})\right\}$,
which depends on the precise location of centre and symmetry properties of 
localized orbitals in the auxiliary subspace. A quantitative idea of 
the {\it smoothness} is obtained with the Berry connection matrix $B_\alpha(k)$:
\begin{equation}
B^{\mu \nu}_\alpha(k)=-Im \langle\tilde{u}_{\mu\mathbf{k}}|
\frac{\partial}{\partial k_\alpha}|\tilde{u}_{\nu\mathbf{k}}\rangle,
\label{phase1}
\end{equation}
and its Fourier components.
If the matrix $B_\alpha(k)$ is diagonal and its diagonal entries are 
constant as a function of $k_\alpha$ (it has vanishing Fourier components for $\mathbf{R} \ne 0$), 
$\Phi_\mu$ are maximally localized in $\alpha$ direction\cite{Sgi, BW}. 
For the systems studied in this work, diagonal elements of $B$ 
are found to be constant as a function of $k$, indicating good localization properties of
$\Phi_\mu$. 

We note that functions $|\Phi_\mu (\mathbf{R})\rangle$ are {\it not} maximally
localized\cite{MV} by construction. However, if desired, the MLWFs can be readily obtained through
a single step of maximal joint diagonalization\cite{cardozo} of the three components of 
position operators in $\Phi$-basis: 
\begin{equation}
\langle r_\alpha\rangle_{\mu,\mu',R_1,R_2}= \int_{\Omega N_{k\alpha}}
  \Phi^\star_\mu (\mathbf{R_1, r})  r_\alpha
 \Phi_{\mu'} (\mathbf{R_2, r}) d \mathbf{r},
\end{equation}
Electronic part of electric polarization is obtained using:
\begin{equation}
P^{el}_\alpha = \frac{q}{\Omega}\sum_\mu \langle \Phi_\mu (\mathbf{R=0}) | r_\alpha 
| \Phi_\mu (\mathbf{R=0}) \rangle.
\label{pol}
\end{equation}
Inclusion of occupation numbers in Eqn \ref{newfns} ensures a simple and general form
for density matrix $\rho(r,r')=\sum_{R,\mu} \langle r| \Phi_\mu(R) \rangle \langle \Phi_\mu(R)|r'
\rangle$.
As the present analysis at {\bf k} is decoupled from that at other {\bf k}'s,
it applies equally well to simple, compound and entangled bands, and can be 
performed efficiently on a parallel computer.
For molecules or confined systems treated within periodic boundary conditions, 
it can be readily applied using a single $\mathbf{k}=(000)$ point; it equally
well applies to confined systems treated with open boundary conditions.
In the present scheme, construction of atom-centered orbitals\cite{kosov} for
covalent systems necessiates expansion of the physical subspace to include anti-bonding
bands. It is somewhat similar to physically intuitive derivation of Wannier functions in 
Ref.\onlinecite{satpathy}.

We illustrate our method within density functional theory through applications to
insulators, metals and molecules.
The energy eigen states, which are the main inputs to our method, are calculated using
ABINIT implementation\cite{abinit} of density functional theory and norm-conserving
pseudopotentials. We use Monkhorst Pack meshes (finer than a 9$\times$9$\times$9 mesh) of 
k-points to sample Brillouin zones. Isosurfaces of localized orbitals (indicating 
isovalue as percentage of its maximum in figures) have been generated using a 
visualization software XcrysDen\cite{xcrysden}.
\begin{figure}[t]
\includegraphics[scale=0.50]{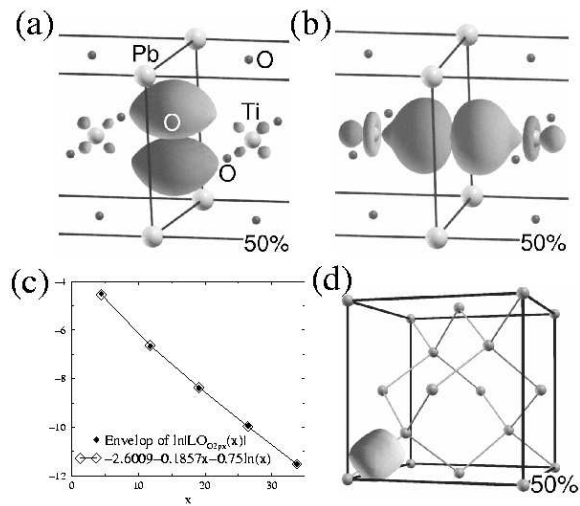} 
\caption{
(a) and (b) shows  2$p_y$ and 2$p_x$ orbitals respectively of a bridging oxygen in PbTiO$_3$. 
(c) shows the nature of decay of the 2$p_x$ orbital shown in (b). 
(d) shows a bond centered orbital representing the Si-Si $\sigma$-bond in bulk Si.  }
\label{insulators}
\end{figure}

For PbTiO$_3$ in the cubic perovskite structure, symmetry of occupied bands implies the
choice of atom-centered orbitals: Pb-centered ($s-$ and $d-$like), Ti-centered ($s-$ and $p-$like)
and O-centered ($s-$ and $p-$like).
Oxygen centered localized orbitals with $p$ symmetry form two groups: (a) ones perpendicular
to the -O-Ti-O- chain (Fig.\ref{insulators}.(a)), (b) ones along the -O-Ti-O- chain
(Fig.\ref{insulators}.(b)). A former has $\pi-$like overlap with $d_{xy}$ orbital of Ti
and a variance of 1.44 \AA$^2$, while the latter has $\sigma-$like overlap 
with $d_{x^2}$ orbital of Ti and a smaller variance of 1.14 \AA$^2$.
For an O-centered orbital with $s$ symmetry, we get a variance of 
0.54 \AA$^2$ (compared with 0.52 \AA$^2$ of Ref. \cite{Gonze-loc}).
We determined Born effective charge ($Z^\star$) of Ti 
using $Z^\star=\Omega \Delta P_\alpha /\Delta d_\alpha$, where  
$\Delta d_\alpha$ is a small displacement of Ti atom that changes the net polarization 
($P_\alpha = P^{ion}_\alpha + P^{el}_\alpha $) by $\Delta P_\alpha$. Our estimate of 
$Z^\star$(Ti)$=$ 7.02 agrees well with a linear response calculation\cite{abinit}, with
contribution from each O-centered $p$ orbital in group (a) and (b) of 1.6 and 0.8 respectively.
Our estimate of $Z^\star$(Pb)$=$3.86, also in good agreement with the linear response result, 
has a contribution of 1.4 from O-centered orbitals in Ti-O planes.
Spatial decay of an O-centered orbital in group (b), shown in Fig. \ref{insulators}(c),
exhibits a power law times times exponential form\cite{kohn73}. The power-law exponent obtained from
a fit is $-0.75$, consistent with a theoretical prediction\cite{HV}.
We confirmed that the diagonal elements of the Berry connection matrix (Eqn.\ref{phase1} 
do not change with $\mathbf{k}$, indicating individually maximal localization of these
orbitals. 
 
For Si in diamond structure, a covalent semiconductor, we illustrate construction of 
bond and atom centered orbitals corresponding to different choices of subspaces.
Symmetries of Bloch states in the occupied subspace (D=4) dictate a choice of bond-centered
orbital with full site symmetry. Corresponding bond centered orbitals (shown in 
Fig.\ref{insulators}.(d)) have a variance  of
2.2 \AA$^2$ (compared with 2.05 \AA$^2$ of an MLWF\cite{MV}) each. 
We find that the diagonal elements of 
Berry connection matrix are indeed constant as a function of $k$, implying maximal
localization of an orbital individually. In the construction of atom-centered
orbitals we use double the number of bands in physical subspace, and atom-centered 
orbitals with $s$ and $p$ symmetry. Our scheme provides two options: (a) treat silicon 
as a metal (using occupation numbers in Eqn.\ref{newfns}), which generates nonorthonormal 
atom-centered orbitals which exactly span the occupied subspace, and (b) not use occupation
numbers, which generates orthonormal atom-centered orbitals that span the occupied
as well as unoccupied states. The former can be important in studies of bonding,
whereas the latter could be useful as basis in $O(N)$ methodology. 
\begin{figure}[b]
\includegraphics[scale=0.40]{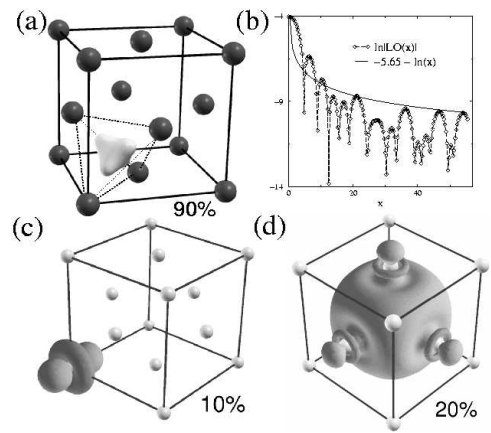} 
\caption{Localized orbitals of metals: an orbital centered at the tetrahedral site of Al (a), 
and its spatial decay (b); atom centered LO with 3$d_{z^2}$ symmetry (c) and an LO
centered at the octahedral site of Cu (d).}
\label{metal}
\end{figure}

We illustrate localized orbitals in metallic systems characterized by multi-centered bonding,
{\it ie.} sharing of electrons among more than two atoms, with examples of Cu and Al in the 
FCC crystal structure. 
Due to partial occupancy of bands in metals, the dimensionality of auxiliary subspace
is taken to be greater than largest number of occupied bands at any $\vec k$, the same 
as the number of energy bands calculated in DFT calculation with temperature smearing
used for occupation numbers.
In the case of Al, we used (guided by symmetry and singular values) spherically symmetric 
orbitals centering at the tetrahedral sites
$(1,1,1)\frac{a}{4}$ in addition to atom-centered $s-$like orbital to construct
an auxiliary subspace. From normalization, we find each localized orbital centered at 
the tetrahedral site (Fig.\ref{metal}.(a)) and the atomic site to be occupied with 1.2$e$ and 
0.6$e$ respectively. The former provides $4-$centered directional bonding among the four Al atoms 
equidistant from its center. Its spatial decay (shown in Fig.\ref{metal}.(b)) exhibits a 
power law decay with an exponent -1, consistent with earlier calculations\cite{metal-decay}
in the free electron limit. In the case of Cu, we used atom centered $s-$ and $d-$like
orbitals and an $s-$ like orbital centered at octahedral site $O=(1,1,1)\frac{a}{2}$.
Among atom-centered orbitals, a $d-$like orbital (Fig.\ref{metal}(c)) is 
quite localized and is occupied with 1.86$e$, while the $s-$like orbital is occupied with
0.69$e$. The orbital centered in the octahedral site (shown in Fig.\ref{metal}.(d)) is occupied
with 1.01$e$ providing a $6-$centered bond, and exhibits slight mixing with atomic $d-$orbitals.
Our results clearly reveal stronger 
directional bonding in Al than in Cu, which was also concluded from charge density analysis
in understanding contrasting mechanical behavior of Al and Cu\cite{sydney}.

\begin{figure}[t]
\includegraphics[scale=0.50]{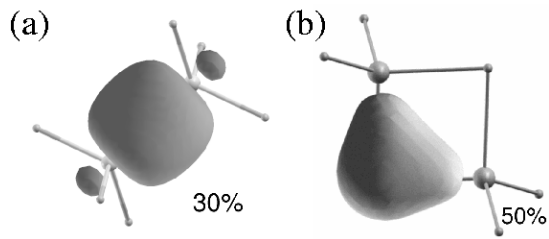} 
\caption{(a) shows a bond centered orbital representing the C-C $\sigma$-bond in C$_2$H$_6$.
(b) shows the B-H-B three center bond in B$_2$H$_6$.   }
\label{molecules}
\end{figure}
Finally, we demonstrate application of the present scheme to localized orbitals in
C$_2$H$_6$ and B$_2$H$_6$ molecules. In the former, we have used bond-centered orbitals
and in the latter, we use bond-centered orbitals for the BH$_2$ radicals and $s-$like 
orbitals centered on H atoms that bridge the two BH$_2$ radicals.
C-C $\sigma-$bonding orbital (Fig.\ref{molecules}(a)) has a variance of 0.83 \AA$^2$, 
which reduces to 0.76 \AA$^2$ after joint diagonalization to obtain an MLWF.
Orbitals corresponding to heteropolar C-H $\sigma$ bonds have their
centroids closer to the H atoms. 
The orbital centered on a bridging H atom of B$_2$H$_6$
(shown in Fig.\ref{molecules}(b)) reveals the well-known 
three-centered bond, and has a variance of 0.96 \AA$^2$ each. 

In conclusion, we have presented a simple method for construction of 
well localized orbitals, that is applicable to extended metals and insulators
as well as finite systems. While demonstrated here for electronic problems, it
can be readily used in treatment of phonons. It should be useful for a 
large number of problems in condensed matter physics.

JB thanks CSIR, India for a research fellowship.
UVW acknowledges support from the DuPont Young Faculty Award and the central computing
facility at JNCASR, funded by the DST, Government of India.

\end{document}